\documentclass{article}

\usepackage{arxiv}

\usepackage[utf8]{inputenc} 
\usepackage[T1]{fontenc}    
\usepackage{hyperref}       
\usepackage{url}            
\usepackage{booktabs}       
\usepackage{amsfonts}       
\usepackage{nicefrac}       
\usepackage{microtype}      
\usepackage{lipsum}
\usepackage{authblk}
\usepackage{graphicx}

\title{Preliminary Report on the Study of Beam-Induced Background Effects at a Muon Collider}

\author[1] {Nazar Bartosik}
\author[2] {Alessandro Bertolin}
\author[3] {Massimo Casarsa}
\author[4] {Francesco Collamati}
\author[5] {Alfredo Ferrari}
\author[8] {Anna Ferrari}
\author[2] {Alessio Gianelle}
\author[6] {Donatella Lucchesi}
\author[9] {Nikolai Mokhov}
\author[8] {Stefan Mueller}
\author[1] {Nadia Pastrone}
\author[7] {Paola Sala}
\author[2] {Lorenzo Sestini}
\author[9] {Sergei Striganov}

  \affil[1]{INFN Sezione di Torino, Torino, Italy}
  \affil[2]{INFN Sezione di Padova , Padova, Italy}
  \affil[3]{INFN Sezione di Trieste , Trieste, Italy}
  \affil[4]{INFN Sezione di Roma , Roma, Italy}
  \affil[5]{CERN, Geneva, Switzerland}
  \affil[6]{University of Padova and INFN Sezione di Padova , Padova, Italy}
  \affil[7]{INFN Sezione di Milano, Milano, Italy}
  \affil[8]{HZDR, Dresden, Germany}
  \affil[9]{Fermilab, Batavia, Illinois, U.S.A}
\begin{document}
\maketitle

\begin{abstract}
Physics at a multi-TeV muon collider needs a change of perspective for the detector design due to the large amount of background induced by muon beam decays. Preliminary studies, based on simulated data, on the composition and the characteristics of the particles originated from the muon decays and reaching the detectors are presented here. The reconstruction performance of the physics processes $H\to b\bar b$ and $Z\to b\bar b$ has been investigated for the time being without the effect of the machine induced background. A preliminary study of the environment hazard due to the radiation induced by neutrino interactions with the matter is presented using the FLUKA simulation program.
\end{abstract}

\section{Introduction}
The quest for higher energy colliders has re-opened the discussion on the possibility to exploit muon collisions to reach multi-TeV energies in the center of mass. During 2018, in preparation for the update of the European strategy for particle physics, the muon collider (MC) working group has submitted an input document~\cite{ESInput} which summarizes the status of different projects. While machines based on different technologies for the muon production have been studied in the past, as presented in Ref.~\cite{ESInput}, the effects of the background induced by the muon beams decays on the physics reaches have not been studied in details, due to the complexity of the beam background at the interaction region (IR). \\
In fact, the muon decays products can arrive to IR from a distance that varies with the beam energy, therefore the collider optics and its superconducting magnets with appropriate protective elements need to be designed and included in simulations~\cite{M1} to evaluate it. Since the level of the background is too high to operate a particle physics detector, two tungsten cone-shaped shields have been proposed, as presented~\cite{M2} and optimized in~\cite{mokhov15,mokhovHF}, to protect the IR and the detector. The exact design of the machine-detector interface (MDI), which includes these shields, is needed to evaluate the distribution of the induced background in any position of the detector. The MAP collaboration~\cite{mapc} studied in details the beam induced background up to the particles distribution on the detectors, then, due to the ending of the research program, no further studies were performed.
The study presented in this paper starts from the latest results obtained by the MAP collaboration, and makes use of their IR and MDI optimized for 1.5-TeV center of mass energy~\cite{mokhov15}. The software framework used for the propagation of the beam background through the detector is the same as that used by MAP before the shutdown of the program.
Recently, a different method to produce muons has been proposed~\cite{ESInput} where the beam intensity is expected to be lower by one or two orders of magnitude than the proton-driven one. In this configuration, the level of the machine-induced background in the detector will be reduced, but at the moment no MDI design is available. Hence, the current studies are performed using the MAP parameters, which in any case represent a worse background scenario.
\section{Beam-induced background simulation}
\label{sec:bckmod}
The composition and the characteristics of the beam-induced background in a muon collider have been studied in detail in Refs.~\cite{mokhov15} and \cite{mokhovHF} for $\mu^+\mu^-$ collisions at $\sqrt{s} = 1.5$ TeV and $\sqrt{s} = 125$ GeV, respectively. Here the most relevant features of the background are summarized.

\begin{figure}[htb]
  \centering
    \includegraphics[width=0.6\textwidth]{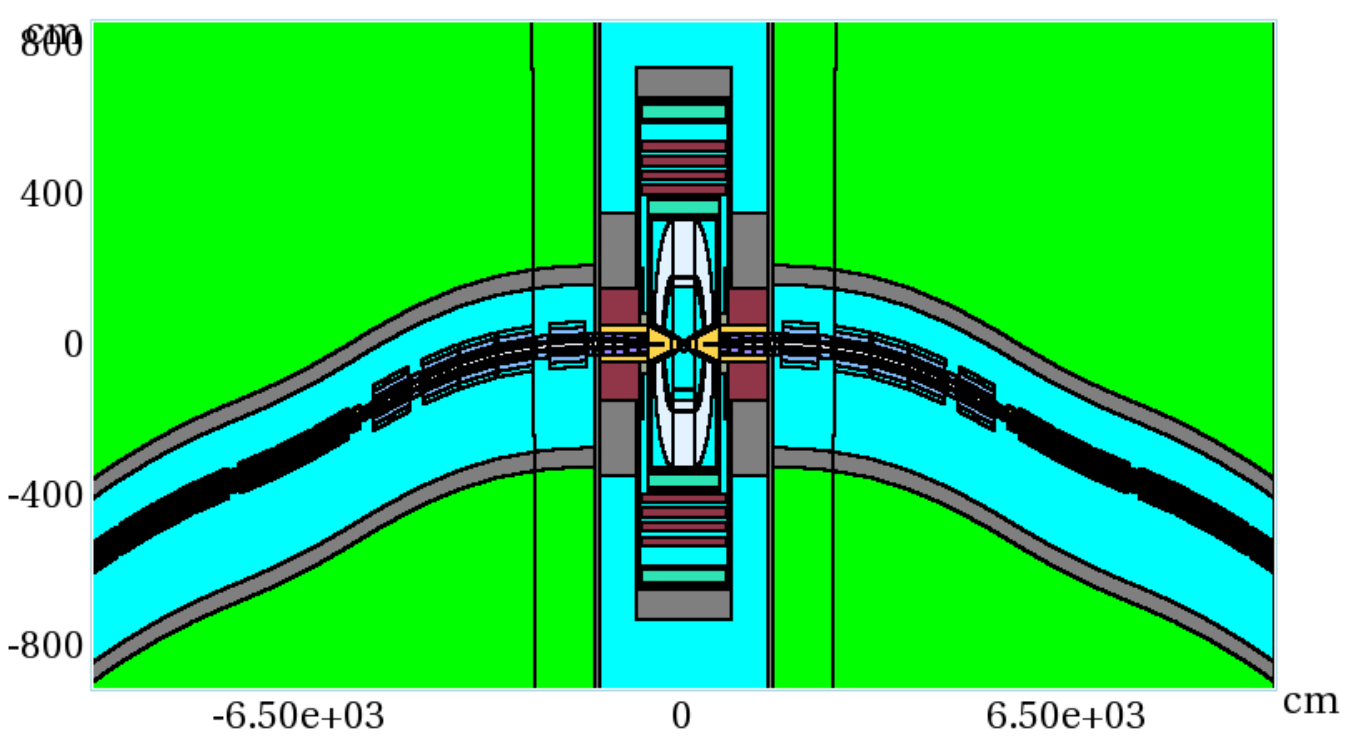}
    \caption{Illustration of the model built for the MARS15 simulation in a range of $\pm$100 m around the interaction point. It includes the machine components in the tunnel and the ILC 4th concept detector~\cite{M3} with the CMS-type tracker upgraded for the High-Luminosity LHC phase. The shielding nozzles are represented in yellow inside the detector. This figure has been reproduced from Ref.~\cite{mokhov15}.
    \label{fig:mars_model}}
\end{figure}
The above-mentioned studies are based on the MARS15 software~\cite{MARS15}, which provides a realistic simulation of the beam-induced background inside the detector.
MARS15 implements a model of the machine-detector interface, the experimental hall, and the machine tunnel with all the collider components in a $\pm$200~m range around the interaction point (IP), including a realistic description of the geometry, the material distribution and the magnetic fields of the lattice elements (see Figure~\ref{fig:mars_model}).
Source of the beam-induced background are the electrons and positrons, generated in muon decays, and the synchrotron photons, successively radiated by the primary $e^\pm$, which interact with the machine components and the surrounding environment producing secondary particles (charged and neutral hadrons, Bethe-Heitler muons, electrons and photons) that eventually may reach the detector.
The actual background level in the detector depends on the beam energy and the configuration of the machine-detector interface. In particular, the MARS15 studies demonstrated that two tungsten cone-shaped shields (nozzles) in proximity of the interaction point, accurately designed and optimized for a specific beam energy, play a crucial role in background mitigation inside the detector. The position and shape of the nozzles are shown in Figs.~\ref{fig:mars_model} and~\ref{fig:det}.

\begin{figure}[t]
  \centering
    \includegraphics[width=0.45\textwidth]{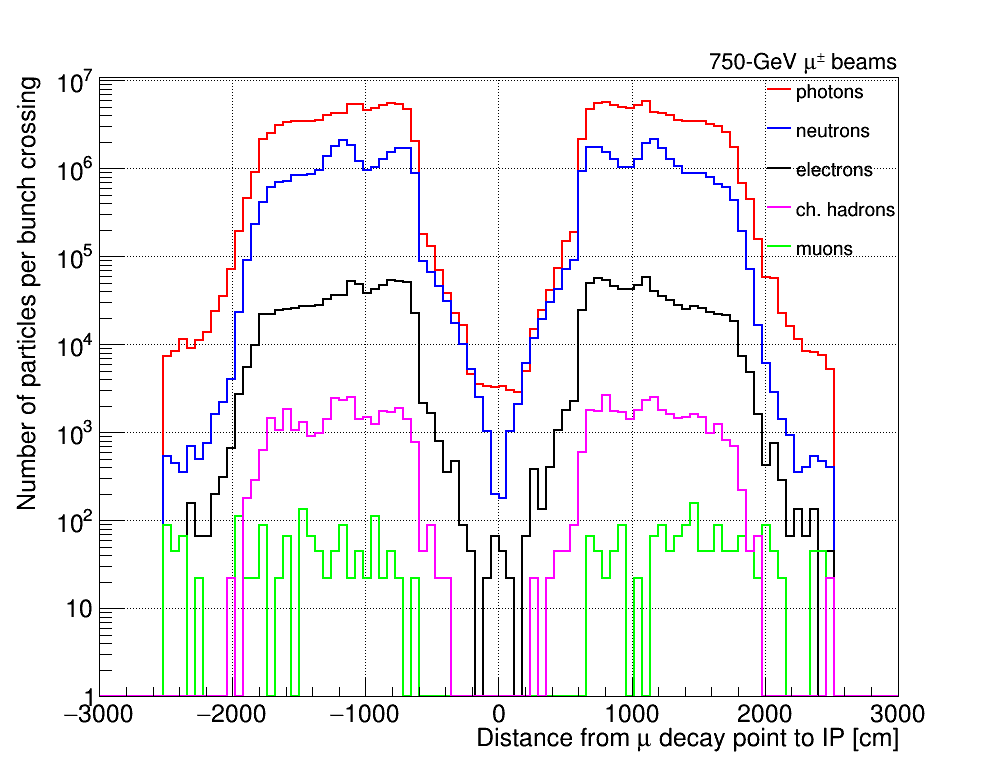}
    \includegraphics[width=0.45\textwidth]{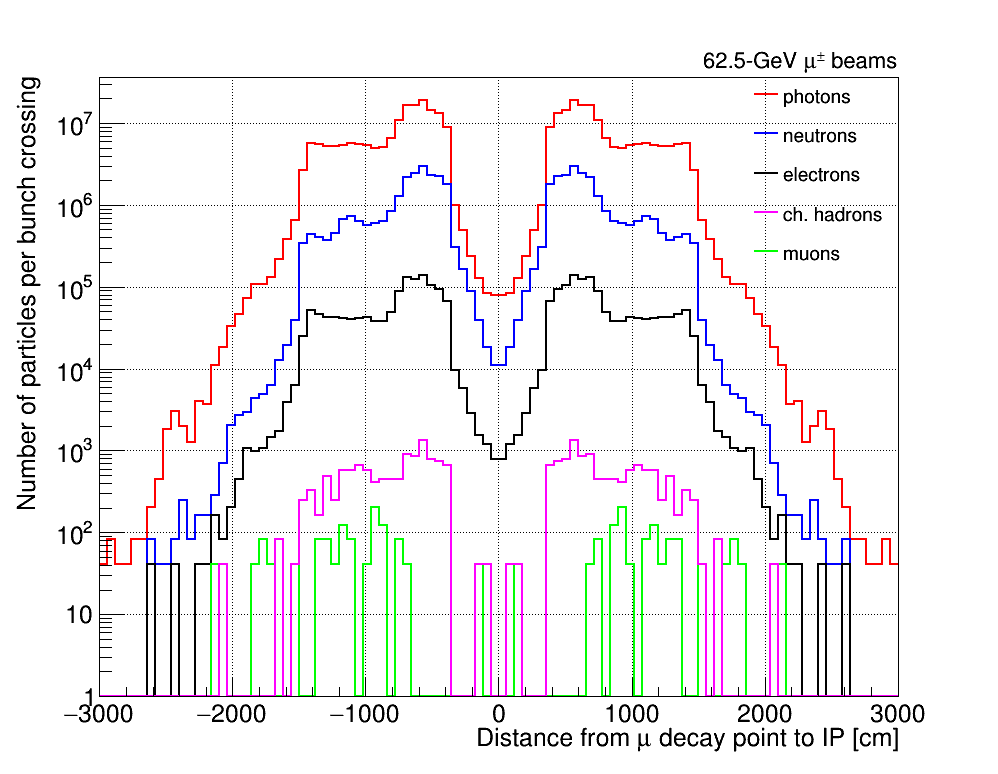}
    \caption{Particle composition of the beam-induced background as a function of the muon decay distance from the interaction point for the cases of a 1.5~TeV (left) and a 125~GeV (right) collider.
    \label{fig:bkg_comp}}
\end{figure}
Figure~\ref{fig:bkg_comp} shows the MARS15-calculated distributions of the different species of background particles as a function of the decay point of the muon from which they had origin. 
The beam-induced background primarily consists of photons, neutrons, $e^\pm$, charge hadrons, and Bethe-Heitler $\mu^\pm$, which are produced by muons decaying in a range of tens of meters around the IP.
Outside that range, which mainly depends on the collider energy and the machine design, the 
detector background contributions become quickly negligible for all components, except from Bethe-Heitler muons, whose range of interest is $\pm$100 m from IP.
This allows to restrict the computationally demanding simulation of the background sample to
muons decaying in a range of $\pm$25 m ($\pm$30 m) around the IP for a 1.5-TeV (125-GeV) collider.
Such a sample accounts for $\sim$80\% of the Bethe-Heitler muons.

\begin{table}[h]
    \centering
    \begin{tabular}{l|cc}
         \hline
         beam energy [GeV]       & 62.5                 & 750                  \\ 
         \hline
         $\mu$ decay length [m]  & $3.9 \times 10^{5}$  & $4.7 \times 10^{6}$ \\
         $\mu$ decays/m per beam & $5.1 \times 10^{6}$ & $4.3 \times 10^{5}$  \\
        \hline
        photons ($E_{\mathrm{ph.}}^{kin} > 0.2$ MeV)            & $3.4 \times 10^8$    & $1.6 \times 10^8$   \\
        neutrons ($E_{\mathrm{n}}^{kin} > 0.1$ MeV)             & $4.6 \times 10^7$     & $4.8 \times 10^7$    \\
        electrons ($E_{\mathrm{el.}}^{kin} > 0.2$ MeV)          & $2.6 \times 10^6$    & $1.5 \times 10^6$   \\
        charged hadrons ($E_{\mathrm{ch. had.}}^{kin} > 1$ MeV) & $2.2 \times 10^4$  & $6.2 \times 10^4$ \\
        muons ($E_{\mathrm{mu.}}^{kin} > 1$ MeV)                & $2.5 \times 10^3$ & $2.7 \times 10^3$  \\
        \hline
    \end{tabular}
    \caption{Expected average number of muon decays per meter and estimated number of background particles entering the detector per bunch crossing for beam energies of 62.5 and 750 GeV. A bunch intensity of $2\times 10^{12}$ is assumed. In parentheses are shown the thresholds set on the particles kinetic energy.
    \label{tab:bkg_comp}}
\end{table}
Table~\ref{tab:bkg_comp} reports the expected average number of muon decays per meter and the estimated yields of background particles entering the detector per bunch crossing for the two considered beam energies, when a bunch intensity of $2 \times 10^{12}$ muons is assumed. 
Due to the effect of the IP and MDI design, specifically optimized for the two energies, the actual level of the detector background does not scale linearly with the beam energy. The suppression factor of the shielding nozzle, protective inserts inside and masks in between the IR superconducting magnets~\cite{M1} is of the order of $\sim$1/500.
Nevertheless, the absolute flux of particles is still very high and poses a serious challenge for the detector readout and particle reconstruction. 
Another potential approach for reducing the flux of background particles is discussed in Section~\ref{sec:future}.

\begin{figure}[t]
  \centering
    \includegraphics[width=0.32\textwidth]{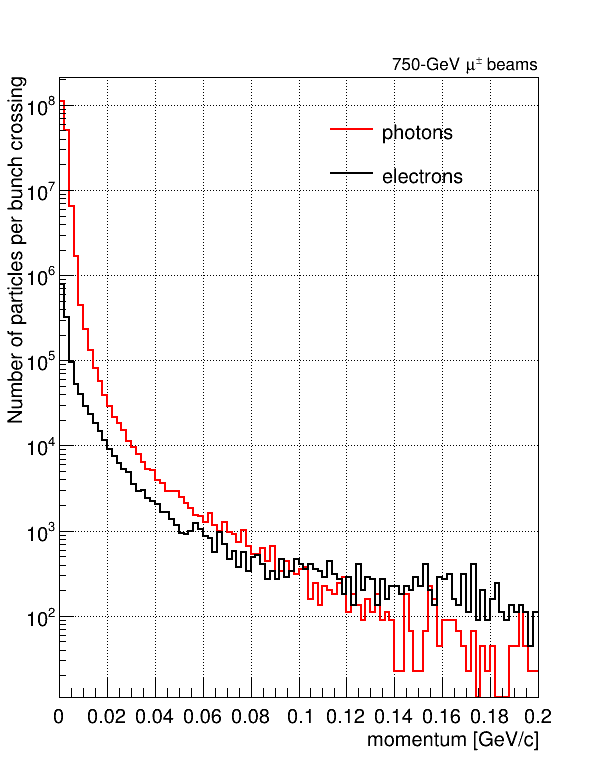}
    \includegraphics[width=0.32\textwidth]{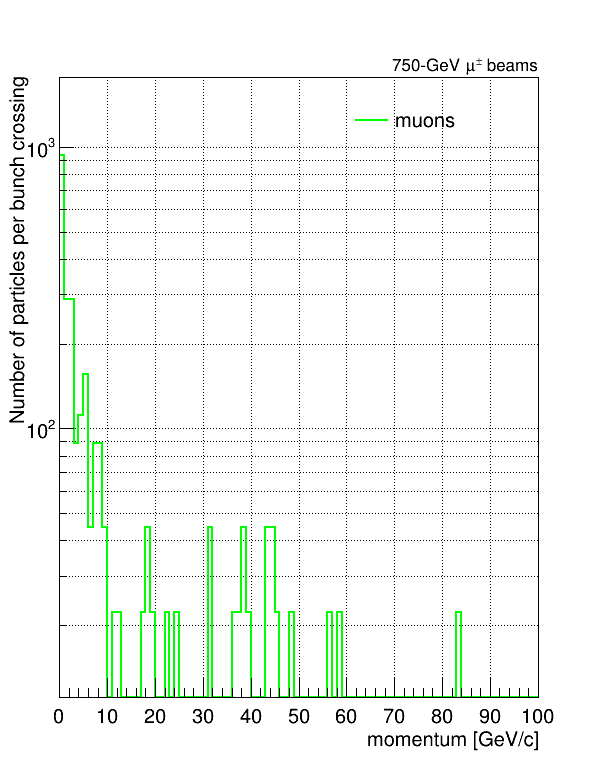}
    \includegraphics[width=0.32\textwidth]{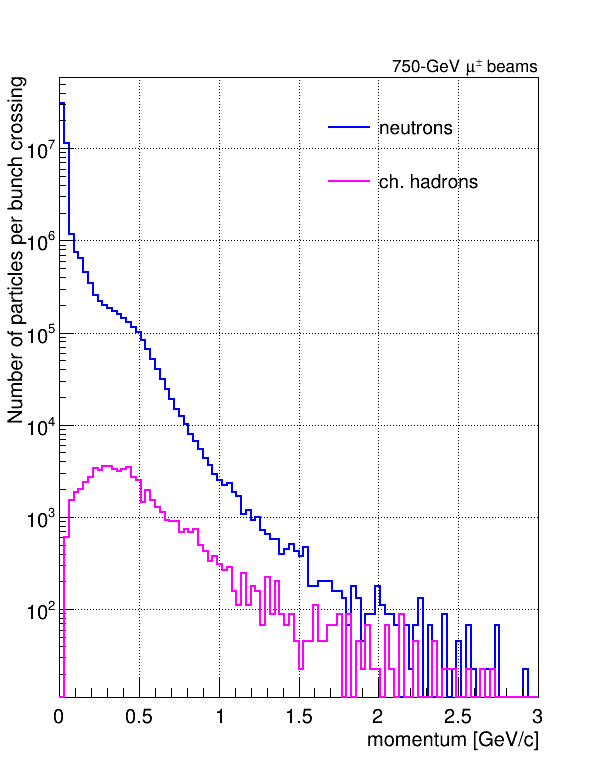}
    \caption{Momentum spectra of the beam-induced background particles at the detector entry point.
    \label{fig:bkg_mom}}
\end{figure}
In Figure~\ref{fig:bkg_mom} the momentum spectra of the beam-induced background are shown for the case of 750-GeV beams. The electromagnetic component presents relatively soft momentum spectra ($\langle p_{\mathrm{ph.}} \rangle = 1.7$ MeV and $\langle p_{\mathrm{el.}} \rangle = 6.4$ MeV), the charged and neutral hadrons have an average momentum of about half a GeV ($\langle p_{\mathrm{n}} \rangle = 477$ MeV and $\langle p_{\mathrm{ch. had.}} \rangle = 481$ MeV), whereas muons momenta are much higher ($\langle p_{\mathrm{mu.}} \rangle = 14$ GeV).

\begin{figure}[t]
  \centering
    \includegraphics[width=0.5\textwidth]{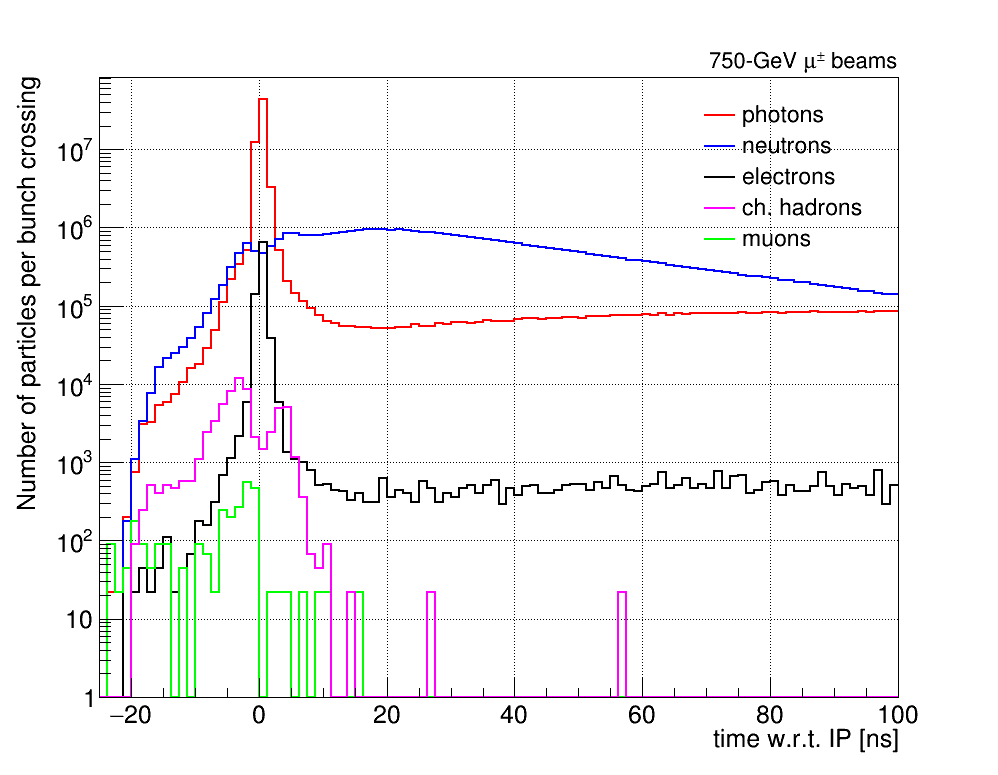}
    \caption{Time of arrival of the background particles at the detector entry point with respect to the interaction point.
    \label{fig:bkg_time}}
\end{figure}
Another distinctive feature of the background particles from muon decays is represented by their timing. Figure~\ref{fig:bkg_time} shows the distributions of the time of arrival at the detector entry point with respect to the bunch crossing time for the different background components.
The evident peaks around zero are due to leakages of mainly photons and electrons in correspondence with the IP, where the shielding is minimal.
\section{Beam-induced background characterization}
\label{sec:bckstudy}
The background samples generated with the MARS15 program are the inputs to the simulation of the detector response in the ILCRoot framework~\cite{vito}. The detector used for the studies presented here has been thought for a MC with a center of mass energy of $1.5$ TeV. Both the framework and the detector are the same as those used by the MAP collaboration before 2014. Several improvements have been achieved since then from the detectors point of view, a new detector design based on up-to-date technologies is needed to compare the physics potential of this machine to the other proposed Future Colliders. The old configuration is used as a starting point for this study, which is going to be updated. In the following, it has to be kept in mind that this is not the best that can be done as of today.

\begin{figure}[ht]
  \centering
\includegraphics{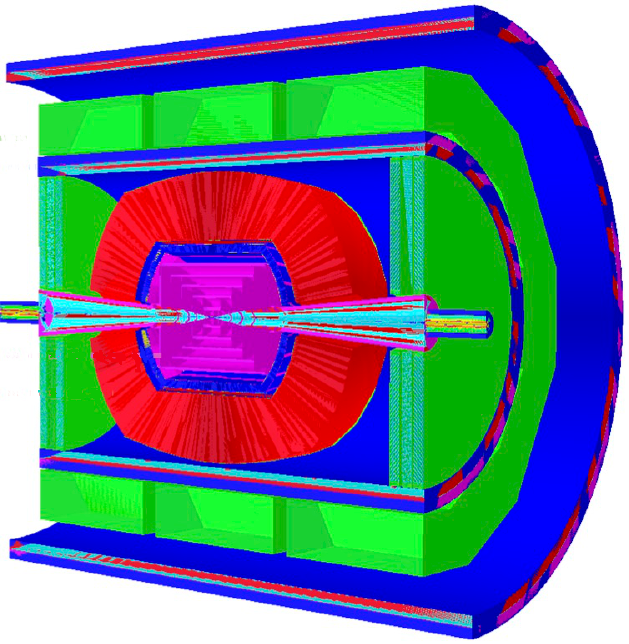}
 \caption{Actual configuration of the detector. From inside to outside, in cyan are the nozzles followed by the tracking system in magenta. The magnetic coil is drawn in blue and the calorimeter system is depicted in red. The muon system, not implemented yet, is represented in green. }
 \label{fig:det}
\end{figure}

The detector simulation includes a vertex (VXD) and a tracking (Tracker) silicon pixel subsystem, as described in Refs.~\cite{vito} and~\cite{ilc4}. Outside a 400-$\mu$m thick Beryllium beam pipe of 2.2-cm radius, the vertex detector covers a region 42-cm long with five cylindrical layers at distances from 3 to 12.9~cm in the transverse plane to the beam axis. The VXD pixel size is $20\time 20~\mu$m. The tracker is constituted by silicon pixel sensors of $50\time 50~\mu$m pitch, mounted on five cylindrical layers from 20 to 120~cm in transverse radius and 330-cm long. The forward region is instrumented with disks also based on silicon pixel sensors, properly shaped in order to host the tungsten shielding nozzles. The full simulation includes electronic noise and thresholds and saturation effects in the final digitized signals. 
The calorimeter is based on a scintillation-Cherenkov dual-readout technique, A Dual-Readout Integrally Active and Non segmented Option (ADRIANO)~\cite{adriano}. The calorimeter simulation for MC in ILCRoot~\cite{adrianosim} considers a fully projective geometry with a polar-angle coverage down to $8.4^o$. The barrel and the endcap regions consist of about 23.6 thousand towers of $1.4^o$ aperture angle of lead glass with scintillating fibers. Cherenkov and scintillation hits are simulated separately and digitized independently. The photodetector noise, wavelength-dependent light attenuation and collection efficiency are taken into account in the simulation of the detector response. Clusters of digitized energy deposits are then used by the jet reconstruction algorithm.\\
The tracking system and the calorimeter are immersed in a solenoidal magnetic field of $3.57~T$.\\
Simulation of the muon detector is not performed given that this is the outermost detector and signatures studied in this article do not include final state muons. Figure~\ref{fig:det} shows a schematic view of the full detector used in the simulation.

\begin{figure}[ht]
  \centering
\includegraphics[width=0.5\textwidth]{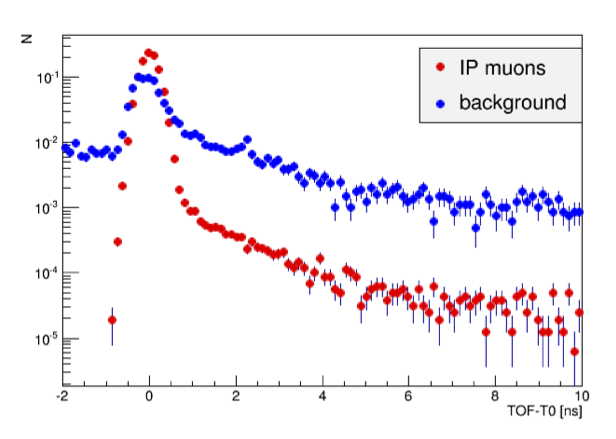}
 \caption{Simulated time of arrival (TOF) of the beam background particles to the tracker modules, summing up all the modules, with respect to the expected time (T0) of a photon emitted from the interaction point and arriving at the same module. 
 }
 \label{fig:timebck}
\end{figure}

\begin{figure}[ht]
  \centering
\includegraphics[width=0.5\textwidth]{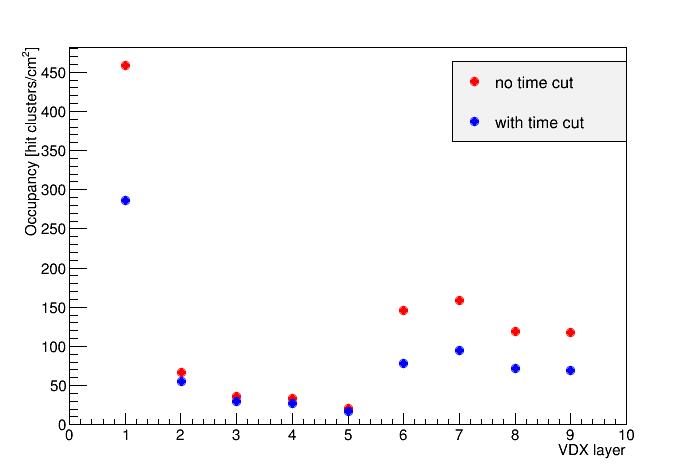}
 \caption{Vertex detector occupancy, defined as the number of hit clusters per $\mathrm{cm}^2$ area, as a function of the detector layers. Layers from 1 to 5 correspond to the barrel layers, from the closer to the more distant from the beam pipe. Layers from 6 to 9 correspond to the endcap layers, from the closer to the more distant from the nominal interaction point. Since endcap layers are on both side with respect to the interaction point, the mean occupancy of left and right layers is shown. Occupancy with and without a time window cut ($\pm 0.5$ ns) is presented.}
 \label{fig:hits}
\end{figure}

Before describing the physical objects reconstruction, we discuss the beam-induced background and the handles available to mitigate its impact. 
As shown in Section~\ref{sec:bckmod}, the noise in the detectors comes from the muon decay products and from their interaction with the nozzles. The spatial and the kinematic distributions show that the tracking system is the most affected detector.
As presented in Ref.~\cite{mokhov15}, the maximum neutron fluence in the innermost layer of the silicon tracker ($R=3$ cm) for a one-year operation is at the level of $10^{8}$~cm$^{-2}$, which is lower than what has been measured for LHC in a similar position and several order of magnitude lower than the $10^{17}$~cm$^{-2}$ expected for FCC-hh~\cite{fcc}.
The number of hits released in the tracking detector by background particles can be reduced by exploiting the time information. As shown in~\cite{vito} and reproduced in this study, these particles have an arrival time distribution that is significantly different from the signal ones. In Figure~\ref{fig:timebck} it is shown the simulated arrival time of particles to the tracker modules with respect to the arrival time of the photons radiated from the interaction point.
By selecting a time window of a few ns around the expected arrival time, a large fraction of the background can be suppressed. This possibility must be studied in detail in the light of the new timing detectors already proposed for HL-LHC where resolutions of tens of picoseconds are achievable~\cite{LHCTimeDet}.
Figure~\ref{fig:hits} shows the hits density as function of the vertex detector layers.
As expected, the first barrel layer, which is closer to the beam, has high hit density, around 450 $\mathrm{cm}^{-2}$ in this configuration. The occupancy of the other barrel layers is significantly lower, at the level or below 50 $\mathrm{cm}^{-2}$, while the endcap layers show an occupancy around 100 $\mathrm{cm}^{-2}$. The cluster density is reduced by applying a time cut, in the first layer it goes down to about 250 $\mathrm{cm}^{-2}$ by requiring a time window of $\pm 0.5$ ns. Improvements are seen also in the endcap layers.
In Ref.~\cite{vito} preliminary studies were presented to illustrate the benefits of using a double layer silicon design. Other strategies, not viable at the time of quoted studies, can be adopted in order to reduce the detector occupancy exploiting the developments done in the latest years for LHC and HL-LHC. This level of background, thought unsustainable years ago, nowadays is comparable with what expected, for example, by the ALICE experiment. In the ALICE Inner Silicon Tracker~\cite{ITSAlice} the hit density of the order of 150 cm$^{-2}$ is foreseen and silicon pixels of $20\times20~\mu$m are adequate to resolve tracks in Pb-Pb events. 
From what discussed above it is obvious that MC detectors will largely benefit from the R\&D planned for the future colliders.

\begin{figure}[ht]
  \centering
\includegraphics[width=0.5\textwidth]{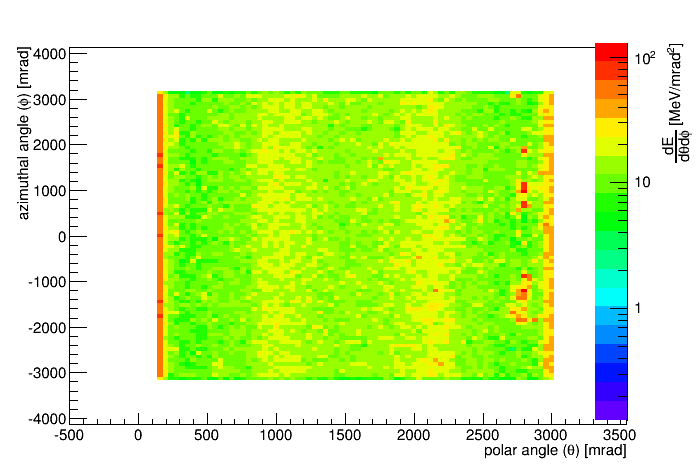}
 \caption{Background energy deposition per bunch crossing in the ADRIANO calorimeter as a function of the polar angle with respect to the beam axis ($\theta$) and the azimuthal angle ($\phi$).}
 \label{fig:adriano_energy}
\end{figure}

The effect of the machine induced background on the calorimeter has been studied and discussed in Ref.~\cite{annacalo}. Figure~\ref{fig:adriano_energy} shows the background energy deposition per bunch crossing in the ADRIANO calorimeter as obtained in this study, which is in agreement with what found before.  As discussed in~\cite{annacalo}, the contamination of this background to the calorimeter clusters associated to signal particles can be reduced by applying appropriate energy thresholds, which is not done here.
\subsection{Physical object reconstruction}
Tracks are reconstructed from clusters of tracker hits that pass the cuts on timing and deposited energy.
The parallel Kalman filter, which is part of the framework, is used for pattern recognition, track propagation and refitting with several track hypotheses in parallel, allowing for cluster sharing between multiple tracks. Both primary tracks (constrained to originate from the interaction point) and secondary tracks (remaining tracks without the constraint) are found with this method.
The performance of the tracking algorithm has been presented in~\cite{annatrack} and was not yet evaluated in this study.

Jet reconstruction was not included in the ILCRoot package, therefore a dedicated algorithm was developed for jet clustering combining information from the tracking and calorimeter detectors. First, the reconstructed tracks and the calorimeter clusters are combined using a Particle Flow (PF) algorithm~\cite{particle_flow}, which performs matching between tracks and clusters to avoid double counting. PF candidates with the transverse momentum greater than 0.5~MeV are then used as input objects in the jet clustering algorithm with the cone size parameter $R=\sqrt{\Delta \eta^2+\Delta \phi^2}$\footnote{$\Delta \phi$ is the difference between the calorimeter cluster and the jet axis in the azimuthal angle. $\Delta \eta$ is the same difference in the pseudo-rapidity variable.} of 2.0 and 1.0 for the 125~GeV and 1.5~TeV cases, respectively. The jet radius is optimized in order to contain most of the energy of $b$-quark jets from the Higgs boson decay. A jet energy correction is applied as a function of the jet transverse momentum. It is determined by comparing the reconstructed jet energy to the energy of jets clustered from Monte Carlo truth-level particles. The jet energy resolution was found to be $11\%$ for the 125~GeV case and $20\%$ at 1.5~TeV, when no beam-induced background is present in the detector.

Jets originating from $b$-quarks are identified using a simple and not yet optimized $b$-tagging algorithm. A secondary vertex, significantly displaced from the primary vertex, formed by at least three tracks is searched. Tracks with an impact parameter greater than 0.04 mm inside the jets are used as inputs to the algorithm. The 2-track vertices are built requiring a distance of closest approach between the two tracks less than 0.02 mm, and a total transverse momentum greater than 2 GeV. Finally, 2-track vertices that share one track are combined to form 3-track vertices.
The $b$-jet tagging efficiency defined as $\epsilon_b=N_{b-tagged}/N_{reconstructed}$ is found to be $\epsilon_b=63\%$ at 125~GeV and  $\epsilon_b=69\%$ at 1.5~TeV. These numbers refer to signal only, since no background is added to the events.

A complete study of tracks efficiency has to be performed including the machine background with a detailed evaluation of the fake tracks. This is mandatory also for the evaluation of the $b$-jet tagging performances in terms of wrong tags.\\
Similar studies have to be completed also for the calorimeter, where anyhow we expect lower contribution from the background.

\section{Characterization of $H\to b\bar b$ and $Z\to b\bar b$ processes}
\label{sec:signals}
\begin{figure}[ht]
\begin{minipage}{0.5\textwidth}
\includegraphics[width=\textwidth]{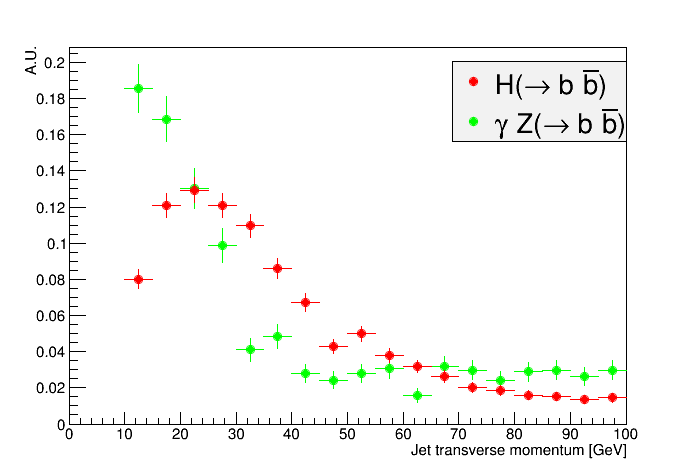}
\end{minipage}
\begin{minipage}{0.5\textwidth}
\includegraphics[width=\textwidth]{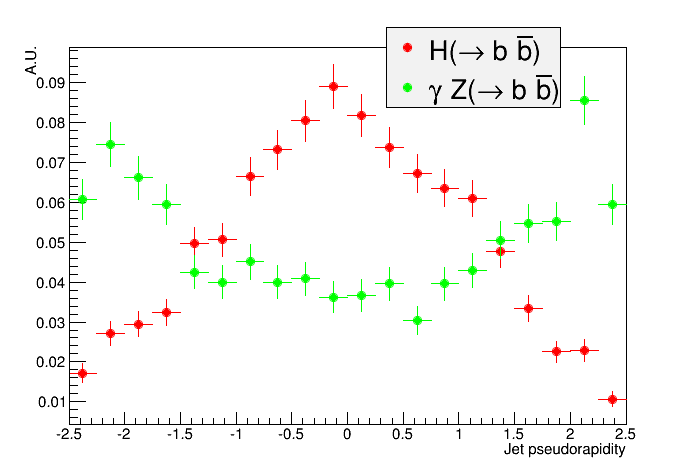}
\end{minipage}
 \caption{Uncorrected jet transverse momentum (left) and jet pseudorapidity (right) in Higgs and $Z$ events produced in 1.5-TeV muon collisions. Higgs and $Z$ distributions are normalized to the same area. Background described in Section~\ref{sec:bckstudy} is not included. }
 \label{fig:higgs_z}
\end{figure}

The reconstruction of $H\to b\bar b$ and $Z\to b\bar b$ is taken as a benchmark to assess the first physics performance of the MC at 1.5 TeV. The two resonances are generated with Pythia 8. In Table~\ref{tab:bb} the production cross sections of processes with two $b$-quarks in the final state are summarized.
\begin{table}[]
    \centering
    \begin{tabular}{l|c}
         \hline
         Process       &        cross section [pb]     \\ 
         \hline
         $\mu^+\mu^-\to \gamma^*/Z\to b \bar b$   & 0.046 \\
         $\mu^+\mu^-\to \gamma^*/Z \gamma^*/Z \to b \bar b$ +X &  0.029 \\
         $\mu^+\mu^-\to \gamma^*/Z \gamma \to b \bar b \gamma$ &  0.12 \\
         $\mu^+\mu^-\to HZ \to b \bar b$ +X  &  0.004   \\
         $\mu^+\mu^-\to \mu^+\mu^-H$ $H \to b \bar b$ (ZZ fusion)  &  0.018   \\
        $\mu^+\mu^-\to \nu _{\mu}\nu_{\mu}H$ $H \to b \bar b$ (WW fusion)  &  0.18   \\
        \hline
    \end{tabular}
    \caption{Cross sections for processes with two $b$-quarks in the final state}.
    \label{tab:bb}
\end{table}
The Higgs and $Z$ signals are generated, simulated and reconstructed following the procedures described above. In this study $b$-tagging is not applied in order to not reduce the statistics, and the background described in Section~\ref{sec:bckstudy} is not included. The fiducial region considered is defined by an uncorrected jet transverse momentum greater than 10 GeV and an absolute jet pseudorapidity lower than 2.5. In Figure~\ref{fig:higgs_z} the uncorrected jet transverse momentum and the jet pseudorapidity in Higgs and $Z$ events are shown. It is evident that jets in Higgs events are well contained in the fiducial region while part of $Z$ events fail the requirements.
In Figure~\ref{fig:higgs_z} the reconstructed di-jet mass distributions for Higgs and $Z$ are shown.  The $Z$ boson is mainly produced in association with a high energy photon (see Table~\ref{tab:bb}), therefore the $Z$ distribution is labeled as $Z+\gamma$.
The relative normalization of the Higgs and $Z$ distributions is taken as the ratio of the expected number of events, considering the selection efficiencies and the cross sections, and it is equal to 12. Although the cross sections are similar, most of the $Z+\gamma$ events fail the fiducial region cuts, therefore a low yield of such events is expected. Since $b$-tagging is not applied a tail at high mass in the $Z$ distribution is present, it corresponds to candidates where the $\gamma$ is reconstructed as a jet.
\begin{figure}[ht]
\begin{minipage}{0.5\textwidth}
\includegraphics[width=\textwidth]{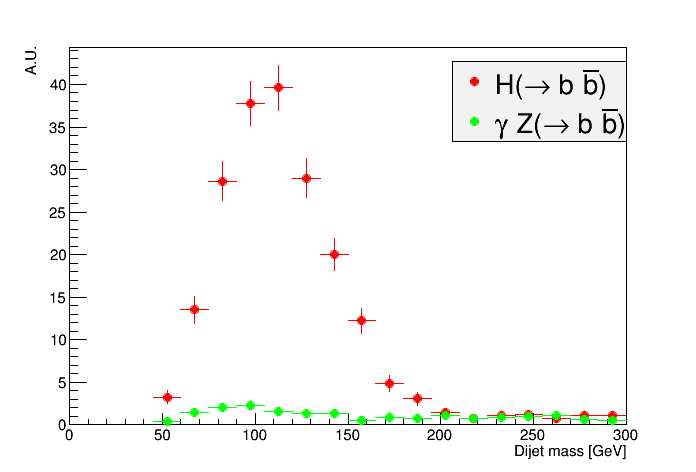}
\end{minipage}
\begin{minipage}{0.5\textwidth}
\includegraphics[width=\textwidth]{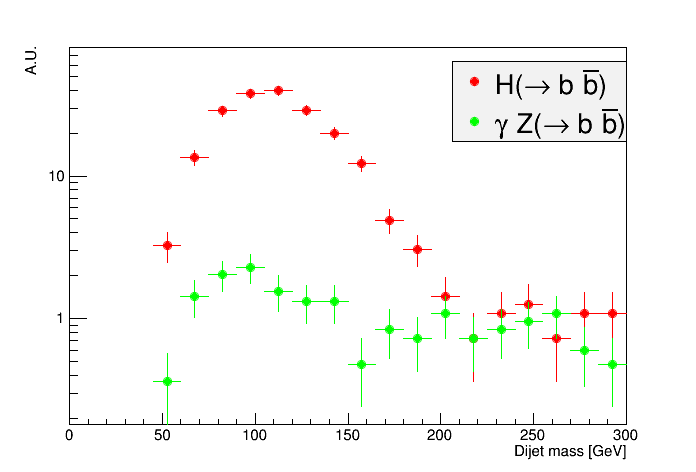}
\end{minipage}
 \caption{Di-jet mass distributions for Higgs and $Z$ produced in 1.5-TeV muon collisions, without and with a logarithmic scale in y-axis (left and right figures, respectively). The relative normalization of the two distributions is equal to the ratio of the expected number of events, considering the selection efficiencies and the cross sections. Background described in Section~\ref{sec:bckstudy} is not included. }
 \label{fig:higgs_z_mass}
\end{figure}

The next step would be to reconstruct the $H\to b \bar b$ and the $Z\to b \bar b$ including the machine-induced background, but unfortunately 
the software and the framework, or at least the knowledge that the authors of this paper have of it, has not allow to do it up to now.
The work is in progress focusing primarly on tracking studies.

\section{Neutrino induced hazard}
The importance of radiation hazard due to highly collimated intense neutrino beams is known since many years.  It has already been studied in an analytic way and with MARS15 simulations, as reported for instance in Refs.~\cite{Mokhov2000,Palmer,King}. 

Concerns come from the dose at the point where the neutrino beam reaches the earth surface, far away from the production point. The dose shall be well below the  recommended annual dose limit for public, presently at 1~mSv/year. A goal of 0.1~mSv/year is assumed here. 
The neutrino beam  spread is roughly given by $1/\gamma$ of the parent muons. At 1 TeV, $1/\gamma \approx 1.\times 10^{-4}$  , resulting in a  100~m spot at a distance of 100 km from the production point. Despite the very small cross section, products from neutrino interactions are concentrated in a small cone, thus delivering a sizable dose. 
When considering a real collider, part of the neutrinos will be produced by muons decaying in the arcs, part in the straight sections. The level and distribution of dose is different in the two situations. In an ideal ring, with no straight sections, the neutrino products will reach the Earth surface along a ring concentric to the collider, at a distance that (for a flat Earth) is roughly proportional to $1/D^2$, were $D$ is the depth at which the collider is situated. The dose from a ring scales approximately with $E^3$, $E$ being the muon energy:  deposited energy scales with $E$, the spot size with $1/\gamma\ \ ~E$, neutrino cross section again with $E$.

Products from straight sections  emerge on a spot-like area, and straight sections dose scales with $E^{4}$  due to an additional $1/\gamma$ factor. 

Dose can be mitigated by proper design limiting straight sections, beam wobbling,  beam focusing/defocusing. Preliminary results shown in the following have to be considered as upper limits.

In view of a full FLUKA~\cite{FLUKA1,FLUKA2} based simulation of detector backgrounds and neutrino hazard in realistic layouts, we describe here the setup and validation of the simulation tools.

\subsection{FLUKA for muon and neutrino transport}
Muon transport in FLUKA includes all interaction processes, from ionization energy losses to bremsstrahlung, pair production, photonuclear interactions and, obviously, decay. Descriptions and comparisons with experimental data are available in the literature, for instance in \cite{Toni,Frascati96}.

The FLUKA neutrino event generator NUNDIS~\cite{NUNDIS} handles quasi-elastic, resonant and deep inelastic neutrino interactions on nucleons and nuclei. The FLUKA nuclear models are exploited to simulate initial and final state effects around neutrino-nucleon interactions. Products of the neutrino interactions can be transported directly in the simulated experimental setup, as was done, for instance, for the ICARUS-T600 experiment in the Gran Sasso underground laboratory\cite{T600} or the ArgoNeut chamber\cite{Argoneutgammas}. In view of the extended energy range foreseen for neutrinos at muon colliders, a check of the NUNDIS prediction at multi-TeV energy has been performed through the comparison with recent IceCube data\cite{IceCube}. NUNDIS results agree with IceCube within experimental errors, showing that the calculated cross section exhibits the correct decrease with respect to linearity versus $\nu$  energy. 

\subsection{Simulation setup}
Simulations described in the following refer to either an idealized ring, assuming continuous bend and no beam divergence, or to idealized straight sections, again with no beam divergence. The Earth surface is assumed to be flat, no mountains, no valleys. A first implementation of wobbling in the ring is also discussed.  
The source, ring or section, is placed at the fixed depth of 550~m. Results for smaller depths can be simply recovered from the depth-distance relation.
Neutrinos are forced to interact along the path from the source to Earth boundary, with a probability proportional to cross section and material density. For the moment, the density and composition of the traversed soil is constant and uniform. Due to the small neutrino cross section, non-uniformity along the path will have no influence on the results. 
Neutrino products are fully transported and ambient dose equivalent (H*(10)) is calculated online trough  convolution of particle fluence and conversion coefficients. 
Results are presented for 1+1 TeV, 1.5+1.5 TeV, 62.5+62.5 GeV, and comparisons are made with previous results from  N. Mokhov and A. Van Ginneken, with related physics models in MARS15, the simulation setup and results of comprehensive simulations described in~\cite{Mokhov2000}
\subsection{Ring}
\begin{figure}[ht]
\begin{minipage}{0.49\textwidth}
\includegraphics[width=\textwidth]{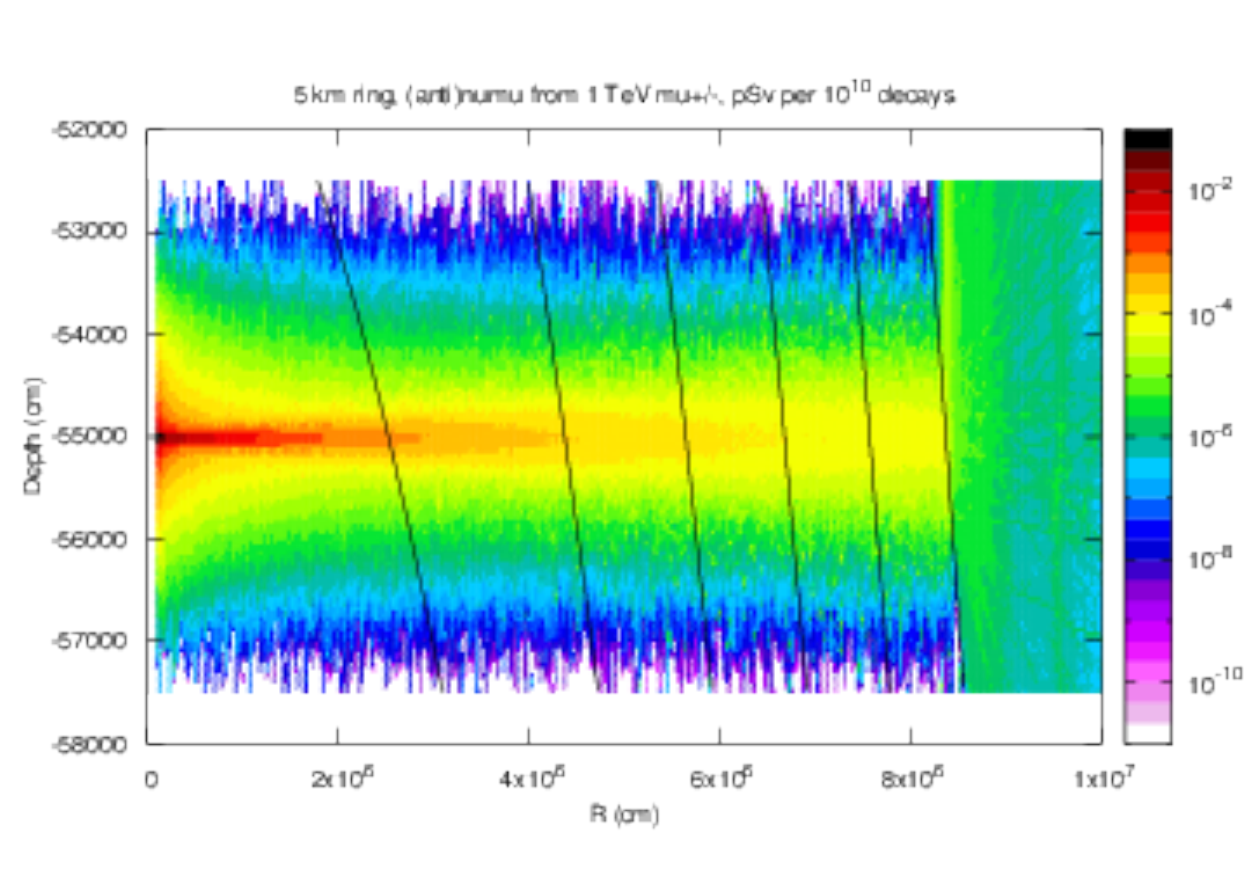}
\end{minipage}
\hfill
\begin{minipage}{0.49\textwidth}
\includegraphics[width=\textwidth]{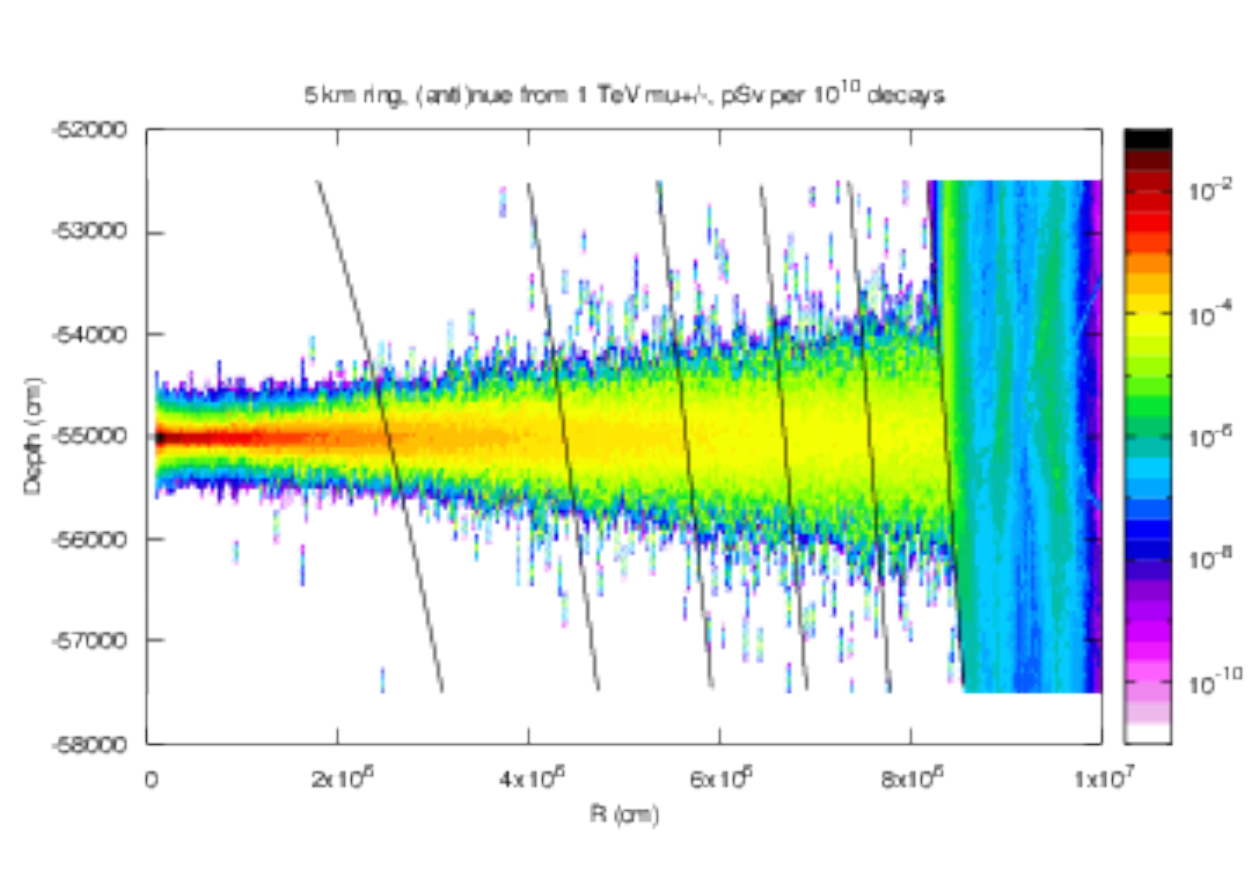}
\end{minipage}
 \caption{ H*(10) from a 1+1TeV ring, versus distance from ring and depth. Color scale units are  pSv/10$^{10}$ muon~decays. Left: from $\nu_{\mu}$ and $\bar\nu_{\mu}$ Right: from $\nu_e$ and $\bar\nu_e$ }\label{fig:ring1tev}
\end{figure}
\begin{figure}[ht]
\begin{minipage}[t]{0.47\textwidth}
\includegraphics[width=\textwidth]{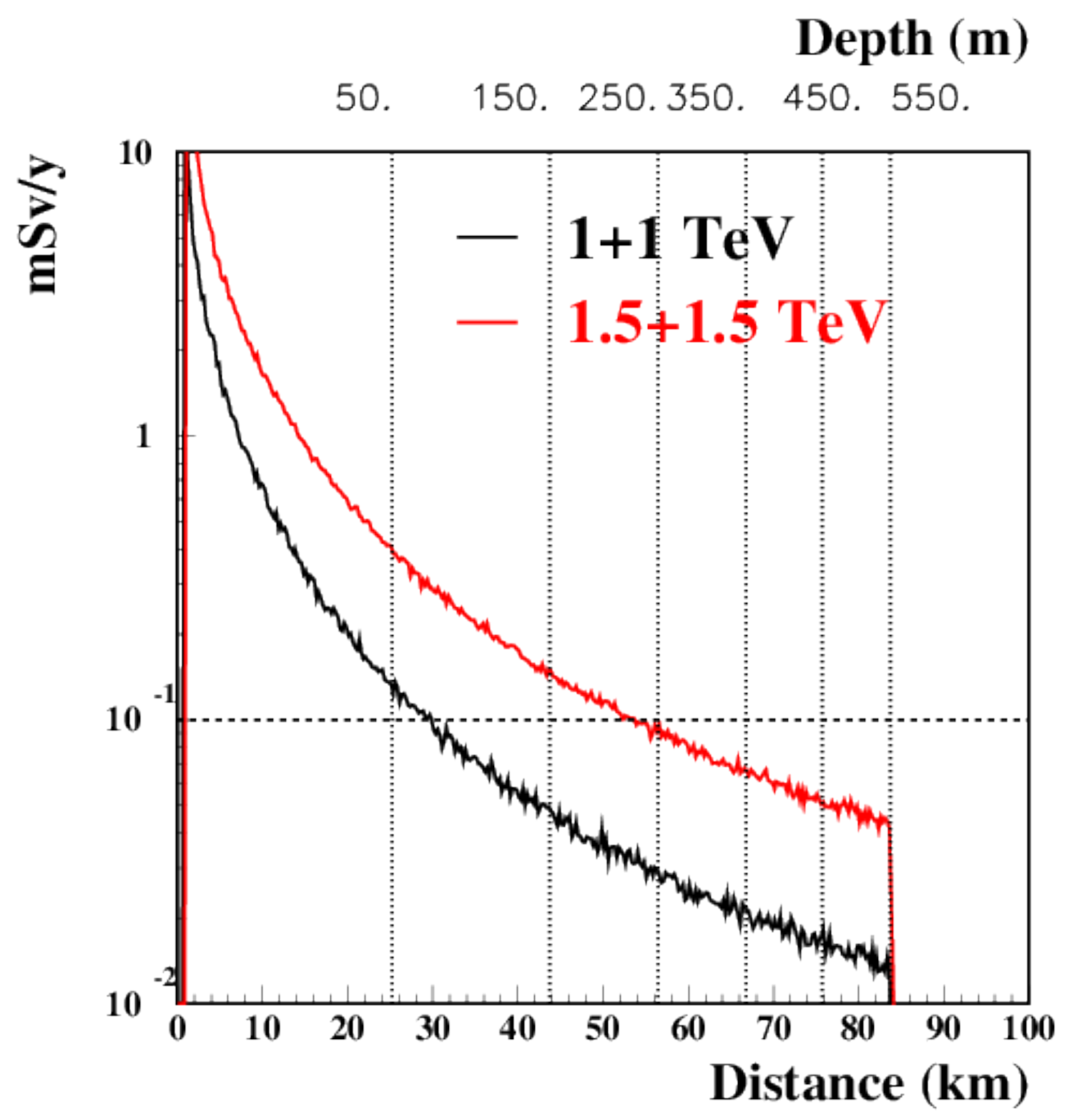}
\caption{H*(10) as a function of distance from ring(bottom axis), or equivalently, depth of the ring (values on top axis, vertical dotted lines) for two muon energies. Averaged over 1m in the vertical plane. Assuming $1.2\times 10^{21}$ decays/year. The dashed horizontal line corresponds to the limit  dose of 0.1~mSv/year}\label{fig:ring1tev1d}
\end{minipage}
\hfill
\begin{minipage}[t]{0.47\textwidth}
\includegraphics[width=\textwidth]{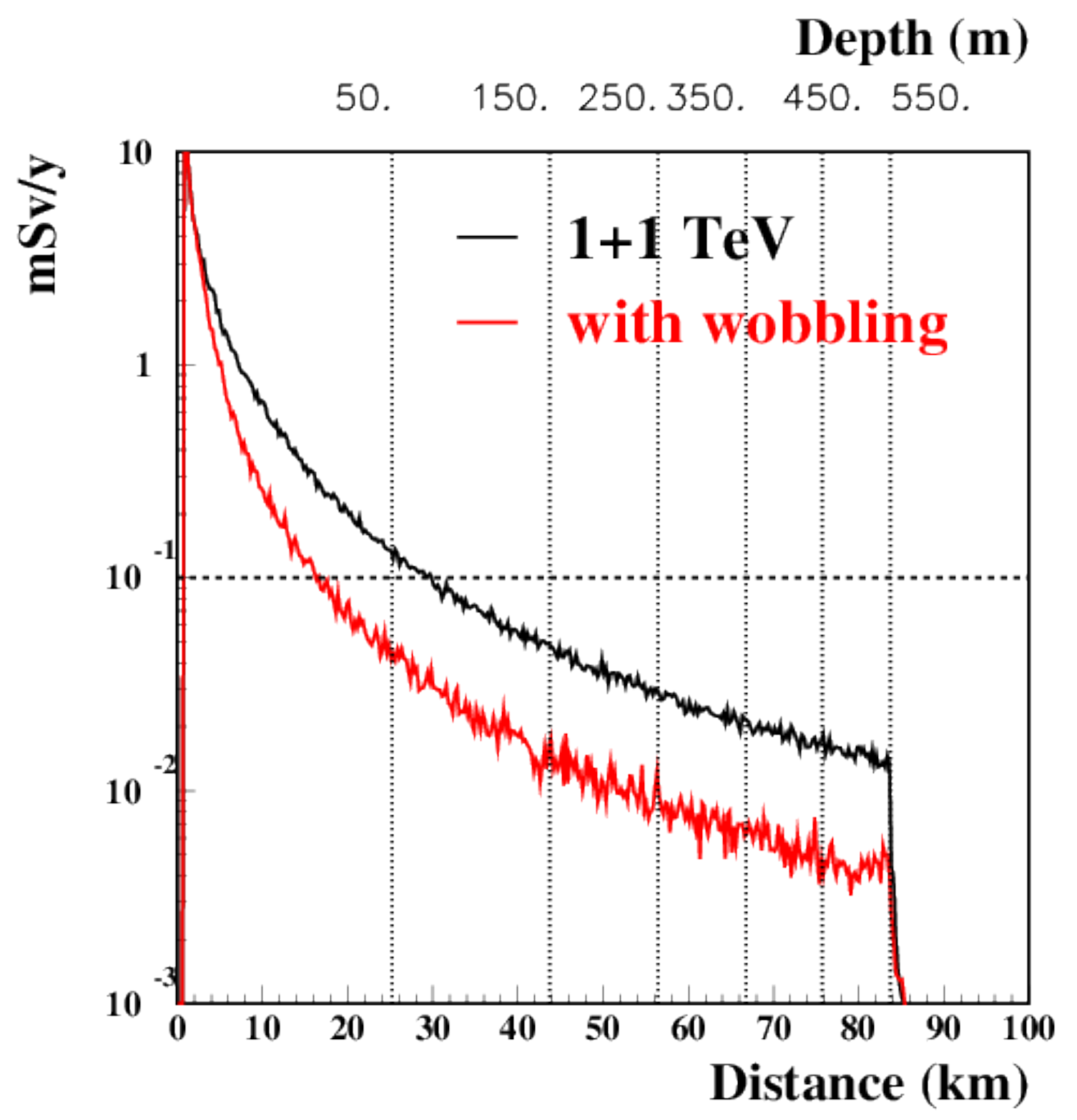}
\caption{As in figure~\protect\ref{fig:ring1tev1d}  for a muon energy of 1~TeV with (red line) and without (black line)  wobbling (see text)} \label{fig:wobbling}
\end{minipage}
\end{figure}

In order to graphically illustrate the shower development, figure \ref{fig:ring1tev} shows the ambient dose equivalent from a ring with circulating $\mu +\bar\mu$ at 1 TeV per beam, as a function of the distance from the ring and of depth.  Curved lines correspond to earth surface for different depth of the ring , in 100 m steps. Even at the maximum distance of about 80~km, from the ring, the shower is vertically contained within $\pm 30$~m. Contribution from electron neutrinos and muon neutrinos are shown separately, to highlight the small difference in vertical spread due to the different ranges of the produced electrons/muons. 
In order to compare with results from MARS15 simulations~\cite{Mokhov2000}, values from figure~\ref{fig:ring1tev} have been normalized to the same number of muon decays per year, namely $1.2\times 10^{21}$. This normalization can correspond  to $2 \times 10^{12} \mu $~/bunch with a frequency of 15~Hz and a run time of 200 days/year.  H*(10) values averaged over 1~m in the vertical plane are shown in figure~\ref{fig:ring1tev1d} for 1 and 1.5 TeV. Present results agree within a factor of 2 with values in figure 8 in~\cite{Mokhov2000}, confirming the soundness of both simulations on this frontier problem.

It appears that dose from a TeV ring is manageable at reasonable collider depths. The approximate $E^3$ scaling is verified, pointing to more problematic situations for multi-TeV colliders. 

It has to be reminded that the values in Figure~\ref{fig:ring1tev1d} refer to a conservative situation, where the muon beam is perfectly parallel. Any variation of beam angular spread along the ring would greatly improve the situation. 

Mitigation procedures have already been put forward, such as vertical wobbling of the beam through inclination of the bending magnets~\cite{Mokhov2000}. Figure~\ref{fig:wobbling} shows the effect of a periodic deflection with a maximum of 100$\mu$rad on the dose from a 1~TeV ring. Reduction of about one order of magnitude can be achieved (again in agreement with MARS15 results~\cite{Mokhov2000}.

\subsection{Straight sections}
\begin{figure}[ht]
\begin{minipage}[t]{0.46\textwidth}
\includegraphics[width=\textwidth]{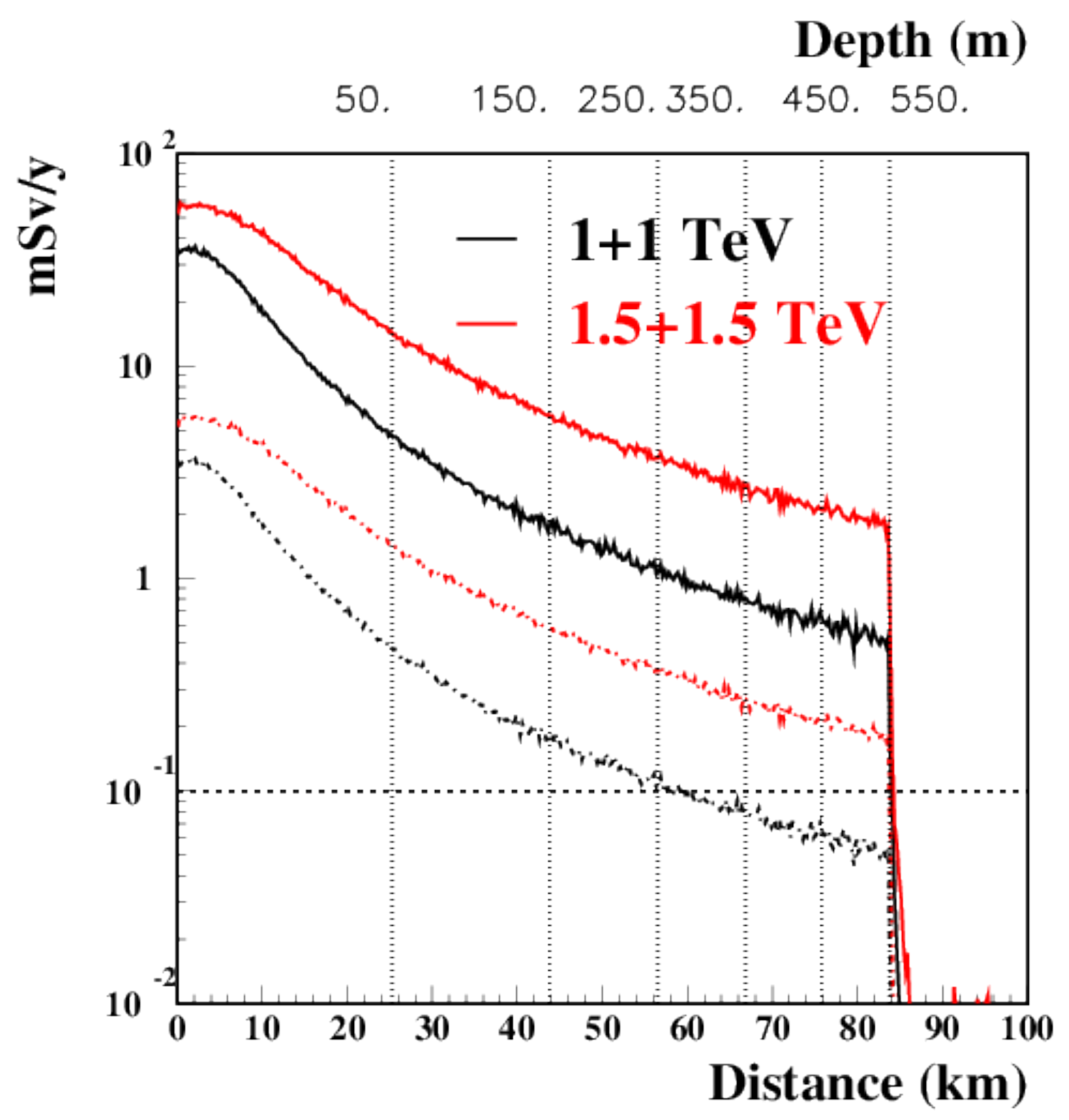}
\end{minipage}
\hfill
\begin{minipage}[t]{0.46\textwidth}
\includegraphics[width=\textwidth]{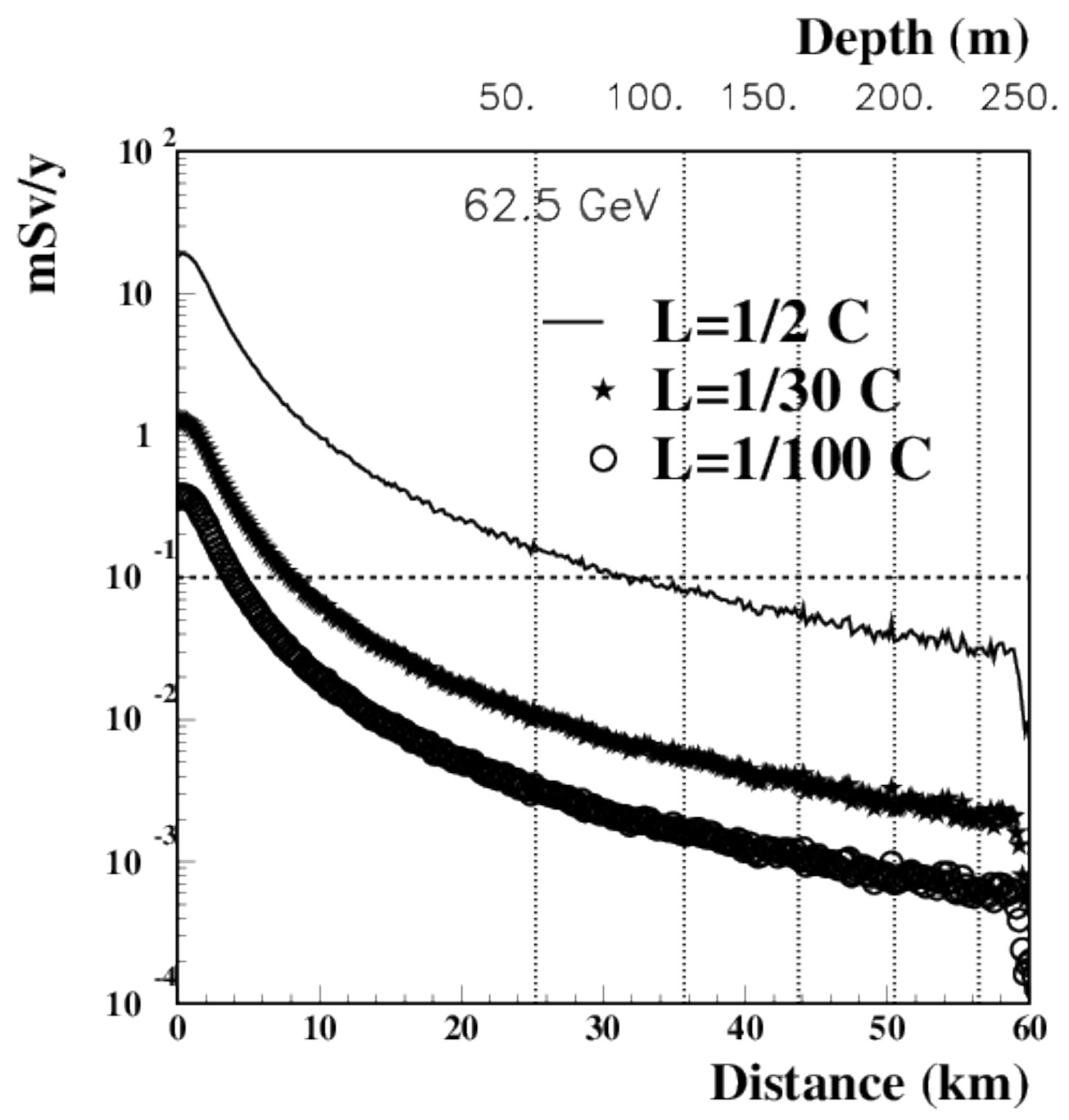}
\end{minipage}
\caption{H*(10) as a function of distance from a straight section (bottom axis), or equivalently, depth of the section (values on top axis, vertical dotted lines). Curves  correspond to different values of  the ratio between section length (L) and total ring length (C). The dashed horizontal line corresponds to the limit  dose of 0.1~mSv/year. Left: for a 1+1  TeV collider, assuming $1.2\times 10^{21}$ decays/year averaged over 1m in the vertical plane. Right: for a 62.5+62.5 GeV collider, assuming $4.8\times 10^{21}$ decays/year averaged over 4m in the vertical plane. For both, the saturation at small distance is due to the too coarse averaging. }
\label{fig:straight}
\end{figure}

Muons and their decay products, from straight sections, are further collimated with respect to arc sections. As a consequence, neutrino-induced dose levels are much higher, and dominate the hazard. Dose depends on the ratio $L/C$ between the length of the considered straight section and the total length of the collider, simply because of the ratio of muon decays. Figure \ref{fig:straight} illustrates the ambient dose equivalent rates for a TeV collider and for a Higgs energy  collider. The latter case is normalized to  $4.8\times 10^{21} \mu$~decays/y, equivalent to $ 4. \times 10^{12} \mu$ /bunch at  30 Hz running  200 days per year.  For a TeV-like collider of total circumference of 10~km, already a straight section of 10~m (L/C=1/1000) produces  doses over 1~mSv /year at the furthest (or deepest) location. The situation is of course much better for low energy rings, except for extreme geometries, not so different however from the ones envisaged for neutrino factories.

Also in this case, beam divergence or focusing/defocusing  would attenuate the risk. Given the size of the neutrino beam, even a minimal spread of the muon beam would largely reduce the dose. Note that the MARS15 results~\cite{mokhov2000} were obtained for the realistic collider optics with beam divergence taken into account. This can explain why the FLUKA results in this application - being rather similar to those from MARS15 - are always higher up to about a factor of two.
\subsection{Summary}
Preliminary simulations of neutrino radiological hazard have been performed with FLUKA. They allowed to validate the tools against past results, and prepare for full simulations in realistic situations. 
These preliminary results show that care must be taken in the design of the machine in order to reduce  neutrino hazard. Limitation in the number and length of straight sections, wobbling, beam divergence, are all factors that will make possible to constrain the dose rate below the one allowed for general public.
\section{Future developments}
\label{sec:future}
These studies are just at the beginning, and there is therefore room for improvement in every aspect. In the following the most urgent ones are described.
\subsection{Framework}
One of the serious limiting factors of this study is the simulation software based on the ILCRoot framework~\cite{vito}. While it has a very detailed implementation of the full simulation of each subdetector, it is rather outdated, suffers from certain performance issues and has a very limited user base.
It is clear that the optimal layout and technologies used in a potential detector at a muon collider would differ from the ones used in this study to benefit from the most recent progress in detector technologies, e.g. high-precision timing at sub-nanosecond level, 3D silicon pixel sensors, high-granularity calorimetry, etc. In view of the complicated arrangement of the code and the absence of other active users, making any significant modifications in the detector design and reconstruction sequence are very time consuming and inefficient. Therefore we foresee a change towards a modern framework that is actively developed and is used for studies on similar projects.

A good candidate could be the simulation and analysis software used by the CLIC experiment~\cite{clic}, which is based on the IlcSoft framework~\cite{ilcsoft}. Besides being a modern set of tools with a large group of active users it has a number of features that are important for the future studies on a muon collider.
One of the important features is the detector description based on the DD4hep toolkit~\cite{dd4hep}, which provides a comprehensive detector definition shared between the simulation, reconstruction and analysis stages, making the iterative process of finding optimal detector configuration much easier.
An almost complete object reconstruction chain is already implemented in the CLIC software, including the novel conformal tracking algorithm~\cite{conformal_tracking}, which minimises the dependence of the code implementation on the actual tracker geometry, Particle Flow, jet reconstruction and b-tagging.
The up-to-date software base makes it easy to parallelise the computationally intensive simulation tasks at the modern cloud-based infrastructures.
\subsection{Machine induced background}
The study presented in this paper is based on background files generated by colleagues of the MAP collaboration, when that program was still active~\cite{MC3tev}.
Despite being this a good starting point, it is however evident that the actual study of a possible muon collider, to be compared to other future accelerators at 3-TeV c.m. energy, must rely on the possibility to have these backgrounds simulated for different machine configurations. It has in fact to be remembered that particles hitting the detector are demonstrated to originate from tens of meters from the IP, and their yield and distribution thus strongly depends on the MDI optics.
In this context, a solid and flexible set of simulation tools must be developed to evaluate the distribution of these background particles and their interaction in the detector as a function of the different optics that will be investigated during the design of the machine.\\
To this aim, the plan is to use FLUKA to simulate the MDI region including the nozzles, in such a way to obtain the distribution of the machine-induced background on the detector surface. These files will be then used as an input for the detailed detector simulation.
A recent comparison of the machine-related detector background at the a 125-GeV muon collider simulated by MARS15 and FLUKA shows a reasonable agreement between the two programs~\cite{MARSvsFLUKA}.
To gain the aforementioned flexibility against possible optics variations, we plan to use scripting tools (e.g. FlukaLineBuilder~\cite{FlukaLineBuilder}) to quickly generate MDI geometries starting from optics files.

To this moment accurately designed nozzles have been extensively studied as the primary way of isolating the detector from the muon-beam decay products. In addition to this passive approach the flux of background particles can be further reduced by an active shielding setup. 
This approach has been facilitated in the SHiP experiment~\cite{ship_shield}, where it allows to reduce the incoming muon flux by $\sim$7 orders of magnitude. A dedicated R\&D on application of this technology to a muon collider configuration would be needed.

\subsection{Detector}
As pointed out in Section~\ref{sec:bckstudy}, the detector simulated with this software package goes back to the time when MAP was active. Nonetheless, it has been of fundamental importance to perform the preliminary studies here presented and in an intermediate stage, before a new framework is adopted, a not-optimal detector can be anyway used to obtain an initial assessment of the Physics performance. Given the peculiarity of the muon collider, a dedicated detector is necessary to fully exploit its Physics potential. This activity will require a dedicated group of experts on different detector technologies.

\section{Conclusions}
We have presented a preliminary study on the effects of the machine-induced background on the detector by using the MAP full simulation framework and MARS15 generated backgrounds for a muon collider with a center of mass energy of 1.5 TeV. The results are in agreement with previously published studies for the tracking system. The calorimeter jet reconstruction, not available in the framework, seems to have similar performance to results presented at conferences. The $b$-jet tagging is studied on events with no machine-induced background, a full evaluation of its performance is in progress. The requirements for the detectors are compared, where available, to those for the detectors at future colliders showing similar challenges. Future studies will largely benefit from the R\&D program foreseen for the detectors for future colliders.
Given the peculiarity of the background, the studies of the physics reaches at different energies need the full simulation which must include the machine-detector-interface. The plan for that include a new full machine-induce background performed with FLUKA and the use of a modern framework.\\
The evaluation of the hazard due to the neutrino interaction with matter has been presented performed with FLUKA in agreement with previous published results obtained with MARS15. FLUKA will be used to evaluate neutrino induced hazard as function of the muon beam energy and interaction region design.

\section*{Acknowledgments}
We owe a huge debt of gratitude to Anna Mazzacane and Vito Di~Benedetto (FNAL) who made this study possible by providing us with the MAP's software framework and simulation code.\\
We would like to thank Mark Palmer (BNL) for his invaluable support and for the advice and the discussions on the studies to be done.
\bibliographystyle{unsrt}  


\end{document}